\documentclass[preprint]{ptephy_v1}%%%%%% to generate preprint number with ptep logo

\preprintnumber{XXXX-XXXX} %%% %%% Insert preprint number here
\usepackage{physics, multirow}
\usepackage{ytableau}
\usepackage{mathtools,booktabs}
\usepackage{nccmath} 
\usepackage{dcolumn,bm}
\usepackage[hyperfootnotes=false]{hyperref}
\hypersetup{
 colorlinks=true,
 citecolor=blue,
 linkcolor=red,
 urlcolor=blue}
\begin{document}

\title{Hidden-Bottom Pentaquarks: Mass Spectrum, Magnetic Moments and Partial Widths}

%%%% To generate auto affiliation numbers please use \author{}\affil{} command

\author{Ankush Sharma, Alka Upadhyay}
\affil{Department of Physics and Material Sciences, Thapar Institute of Engineering and Technology,
Patiala, India \email{ankushsharma2540.as@gmail.com}}

%%% To include the collaborator name... Please use the command "\collaborator"
%%% For example: \collaborator{ATLAS Collaboration}

\begin{abstract}
By taking into light the discovery of pentaquark structures like $P_{\psi s}^\Lambda(4338)^0$, $P_c(4380)$ and $P_c(4450)$, we performed the spectroscopy of hidden-bottom pentaquarks. By utilizing special unitary representations, we systematically classified the hidden bottom pentaquarks into two distinct configurations within the SU(3) flavor representation: the octet and decuplet. In this study, we employed an extended version of the GR mass formula along with the effective mass scheme to provide estimations of the masses associated with hidden-bottom pentaquarks. Furthermore, we extend our analysis to estimate the magnetic moments using the effective mass scheme and screened charge scheme. 
Moreover, by employing the effective Lagrangian, we computed the partial widths for the octet configuration. This comprehensive analysis offers crucial insights into the decay mechanisms and lifetimes of these exotic particles, enhancing our understanding of their fundamental properties. Our findings, which include calculations of masses and magnetic moments, demonstrate reasonable agreement with existing theoretical predictions. 
\end{abstract}
%\vspace{-5.5cm}
\subjectindex{B60}
\maketitle
\section{Introduction}
%\label{sec:intro}
Exotic hadron field has experienced significant expansion in recent times. The LHCb and Belle collaborations have made significant progress in the field of experimental physics, uncovering several exotic states such as tetraquarks and pentaquarks. Quantum chromodynamics (QCD) also acknowledges these entities as bound states and verifies their presence.  The LHCb collaboration has made significant progress in the detection of exotic hadrons. Two singly heavy tetraquark states were observed in 2022 by the LHCb collaboration having quark contents $u\bar{d}c\bar{s}$ and $\bar{u}dc\bar{s}$. These states were observed with a statistical significance of 6.5 $\sigma$ and 8 $\sigma$, respectively \cite{2900}. The experimental mass and width are  $2908\pm 11 \pm 20$ MeV and $\Gamma_{exp}= 136 \pm 23 \pm 13$ MeV respectively.  Additionally, in the $B^-\rightarrow J/\psi\Lambda p$ decays, a hidden-charm pentaquark structure $P_{\psi s}^\Lambda(4338)^0$ ($udsc\bar{c}$) is observed with a statistical significance of 15 $\sigma$ \cite{4338}. The mass and the width of the $P_{\psi s}^\Lambda(4338)^0$ state are  $4338.2\pm0.7\pm0.4$ MeV and $7.0\pm1.2\pm1.3$ MeV, respectively. In 2015, the LHCb collaboration identified two pentaquark structures hidden-charm known as $P_c(4380)$ and $P_c(4450)$ with masses and corresponding widths as
$4380 \pm 8 \pm 29$ MeV and $4449.8 \pm 1.7 \pm 2.5 $ MeV and $205 \pm 18 \pm 86 $ MeV and
$39 \pm 5 \pm 19$ MeV respectively \cite{PhysRevLett.115.072001}. In 2019, a pentaquark state named as $P_c(4312)^+$, observed decaying into $J/\psi p$, with a significance of $7.3\sigma$ in $\Lambda_b^0
 \rightarrow J/\psi p K^-$ decays \cite{PhysRevLett.122.222001}. In 2020, the first confirmation of charm-strange structure in the $J/\psi \lambda$ invariant mass distribution was obtained from $\Xi_b^- \rightarrow J/\psi \lambda K^-$ decays with mass and width as $4458.8 \pm 2.94_{-1.1}^{+4.7}$ MeV and
$17.3 \pm 6.5_{-5.7}^{+8.0}$ MeV, respectively \cite{20211278}. Thus, the field of exotic hadron has grown rapidly and inspired theorists to study their dynamics by applying different frameworks. In this article, a study of masses and magnetic moments of hidden-bottom pentaquarks has been carried out. Several theoretical frameworks have been proposed to analyze the hidden-bottom states such as the Chiral quark model \cite{chiral}, quark delocalization color screening model \cite{color}, vector meson dominance model \cite{vector}, color magnetic interaction \cite{colormagnetic}, QCD sum rules \cite{QCDS} etc. Future experiments could explore the possibility of observing a hidden-bottom pentaquark. The $P_c$ pentaquark states were discovered through the decay of $\Lambda_b$ baryon at the LHCb experiment. However, no such decay process has been observed for the hidden-bottom pentaquarks. Pentaquarks composed of three light quarks and a pair of heavy quarks, can be created by stimulating the nucleon with a photon or pi meson, causing a pair of hidden-bottom quarks to emerge. Thus, in the future, hidden-bottom pentaquarks may be discovered by stimulating the nucleon with photons or mesons. \cite{Production}.\\
By utilizing the SU(3) flavor representation, we studied the classification scheme for hidden-bottom pentaquarks. In Ref.\cite{Santo} and Ref.\cite{Ankush}, hidden-charm pentaquarks are classified in the octet (8) and decuplet (10) configurations with the help of SU(3)  representation. In this work, hidden-bottom pentaquarks are classified in octet and decuplet configurations, and the systematic study of masses, magnetic moments, and partial widths has been carried out. We used an extended GR mass formula and an effective mass scheme to estimate the masses of hidden-bottom pentaquarks. The Gursey-Radicati mass formula appears to be a highly useful tool for analyzing hadron masses. \cite{FG}. The GR mass formula describes the relationship between a particle's mass and its quantum numbers. Several studies have been proposed utilizing an extended version of the GR mass formula to accurately predict the masses of exotic hadrons. \cite{Santo, HOLMA, Sharma_2023, Ankush}. Furthermore, pentaquarks are exotic hadrons composed of four quarks and an antiquark, and the study of magnetic moment is particularly interesting due to their distinctive composition.  Pentaquark magnetic moments provide light on their internal structure and the behavior of their quark constituents and act as a sensitive indicator of the distributions of charges and currents within these particles. We analyze the magnetic moments of hidden-bottom pentaquarks using the effective mass and screened charge scheme. Both effective mass and screened charge schemes are theoretical frameworks employed to simplify complex interactions between particles. Particles in certain physical systems exhibit an effective mass that differs from their intrinsic or rest mass, owing to interactions with their environment. The effective mass is a parameter that is employed to represent the modified mass inside the specific system being studied. In certain situations, charges can be screened or shielded by the presence of other charges or a surrounding medium. The screened charge scheme accounts for this screening effect, modifying the observed charge experienced by a particle in a given environment of a hadronic state. The organization of this work is as follows: The theoretical formalism is briefly described in Section 2. It comprises the effective mass scheme, the screened charge scheme, an extension of the Gursey-Radicati mass formula, and the classification system for hidden-bottom pentaquarks using the SU(3) flavor and SU(2) spin representation. Section 3 describes the octet and decuplet representation of hidden-bottom pentaquarks, Section 4 describes the partial decay widths using the effective Lagrangian method and Section 5 concludes the article summary.

\section{Theoretical Formalism}
\subsection{Classification scheme for Hidden-Bottom Pentaquarks.}
To classify the hidden-bottom pentaquarks, we used special unitary representations. By employing the SU(3) flavor representation, we classified the hidden-bottom pentaquarks into their allowed flavor multiplets. where each quark as well as an antiquark is represented by a fundamental '3' representation. Due to the presence of $b\Bar{b}$ pair in hidden-bottom pentaquarks, the flavor wavefunction for hidden-bottom pentaquarks follows the same classification scheme as that of light baryons ($qqq$). Therefore, allowed multiplets for hidden-bottom pentaquarks are a singlet, two octets, and a decuplet, which can be written as:
 \begin{equation}
     [111]_1,\hspace{0.5cm} 2[21]_8,\hspace{0.5cm}   [3]_{10}
     \label{flavor}
 \end{equation}
we studied the octet and decuplet configurations for hidden-bottom pentaquarks. We employed the SU(2) spin representation, in which a fundamental '2' representation is used to represent both quarks and antiquarks, to describe the spin wave function of the pentaquark.
 Therefore, allowed spin multiplets for pentaquark ($qqqq\Bar{q}$) states are:
\begin{equation}
     [6],\hspace{0.5cm}  4[4],\hspace{0.5cm}    5[2]
     \label{spin}
 \end{equation}
where [6], [4], [2] multiplets belong to spin-5/2, 3/2, and 1/2 respectively. By using the eq. \eqref{flavor} and \eqref{spin} for flavor and spin, we analyze the hidden-bottom pentaquarks into octet and decuplet configurations with spin-3/2 and 5/2 respectively. The spin wavefunction for the spin-5/2 pentaquarks can be expressed as \cite{young}:
\begin{center}
  \begin{ytableau}
    \none & 1 & 2 & 3 & 4 & 5
\end{ytableau}${\chi_1}^P$ 
\end{center}
Similarly, Table \ref{tab:1} displays the spin wavefunction for spin-3/2 pentaquarks using the Young tableau representation.
\begin{table}[h!]
    \centering
    \caption{Representation of spin-3/2 pentaquarks using the Young Tableau technique \cite{young}.}
    \begin{tabular}{cc}
     \begin{ytableau}
    \none & 1 & 2 & 3 & 4 \\
    \none & 5
\end{ytableau}$\chi_2^P$ & \begin{ytableau}
    \none & 1 & 2 & 3 & 5 \\
    \none & 4
\end{ytableau}$\chi_3^P$  \\ &\\
     \begin{ytableau}
    \none & 1 & 2 & 4 & 5 \\
    \none & 3
\end{ytableau}$\chi_4^P$ &
  \begin{ytableau}
    \none & 1 & 3 & 4 & 5 \\
    \none & 2
\end{ytableau}$\chi_5^P$ 
 \end{tabular}
 \label{tab:1}
\end{table}
Spin-5/2 pentaquarks have one spin symmetry, while spin-3/2 pentaquarks have four spin symmetries. Spin symmetries are utilized for computing the magnetic moments of octet and decuplet particles. $\chi_1^P$ symmetry is used for spin-5/2 pentaquarks and  $\chi_2^P$ have been used for spin-3/2 pentaquarks. In the following section, we presented an extended GR mass formula for predicting the masses of hidden-bottom pentaquarks.

\subsection{Extended Gursey-Radicati mass formula}
Extending the Gursey-Radicati mass formula is a valuable method for calculating the masses of exotic hadrons. F. Gursey and L. Radicati initially established the mass formula to analyze the masses of baryons \cite{FG}. Moreover, E. Santopinto and A. Giachino extended it for masses of pentaquarks with hidden charm as \cite{Santo}:
\begin{align}
  M_{GR} = M_0 &+ AS(S+1) + DY  + E[I(I+1) -1/4 Y^2] + G C_2(SU(3)) + F N_c
\end{align}
This mass formula can be used to determine the masses of hadrons that include charm quarks. P. Holma and T. Ohlson expanded this formula to incorporate the influence of the bottom quark \cite{HOLMA}:
 \begin{align}
  M_{GR} = \xi M_0 &+ AS(S+1) + DY  + E[I(I+1)  -1/4 Y^2] + G C_2(SU(3)) + \sum_{i=c,b} F_i N_i
  \label{mass formula}
  \end{align}
Here, $M_0$ represents the scale parameter, whereas $\xi$ denotes the correction factor to the scale parameter, which is associated with the number of quarks comprising the hadron. Each quark contributes 1/3 to the mass of the hadron, resulting in $\xi$ values of 5/3 for pentaquarks. The quantum numbers for spin, isospin, and hypercharge are represented as $S$, $I$, and $Y$, respectively. $N_i$ represents the number of charm and bottom quarks (anti-quarks). $C_2(SU(3))$ is the Casimir operator for SU(3) representation and it has a value of 3 and 6 for the octet and decuplet configuration of SU(3) flavor representation respectively. The mass formula parameters are derived from Ref. \cite{Santo, HOLMA}, where they are determined based on the known baryon spectrum due to limited data on experimentally observed pentaquark states. This set of parameters successfully predicted the exotic hadron masses \cite{Santo, HOLMA, Ankush, Sharma_2023}. Mass formula parameters with their corresponding uncertainties are reported in Table \ref{tab:2}.

\begin{table}[ht]
\centering
\caption{The values of the mass formula parameters used in the extension of the Gursey-Radicati mass formula, together with their respective uncertainties. \cite{Santo}}
\tabcolsep 1.0mm  
\begin{tabular}{cccccccc}
       \hline
       \hline
         & $M_0$ & A & D & E & G & $F_c$ & $F_b$ \\
         \hline
        Value[MeV] & 940.0 & 23.0 & -158.3 & 32.0 & 52.5 & 1354.6  & 4820 \cite{HOLMA} \\
        \hline
        Uncert.[MeV] & 1.5 & 1.2 & 1.3 & 1.3 & 1.3 & 18.2  & 34.4  \\
        \hline
        \hline
       \end{tabular}
        \label{tab:2}
   \end{table}

By using eq. \eqref{mass formula}, the mass spectrum of hidden-bottom pentaquarks in octet and decuplet configuration are reported in Tables \ref{tab:3} and \ref{tab:4} respectively. In the next subsection, the effective mass and screened charge scheme are introduced to calculate the magnetic moment of the hidden-bottom pentaquarks.

\subsection{Effective Mass Scheme}
Studying the magnetic moment of hidden-bottom pentaquarks offers valuable insights into their charge distribution and magnetization, aiding in a deeper comprehension of their geometric structures. This work determined the effective mass of quarks (anti-quarks) by considering their interactions with neighboring quarks through a single gluon exchange mechanism. We determined the masses and magnetic moments of the hidden-bottom pentaquark states using the effective quark masses. There are two ways to express the mass of pentaquarks \cite{VERMA}:
\begin{align}
        M_P =& \sum_{i=1}^5 m_i^{eff} \label{10} \\
   M_P =& \sum_{i=1}^5 m_i + \sum_{i<j} b_{ij} s_i.s_j
    \label{11}
\end{align}
 In this case, $m_i^{eff}$ denotes the effective mass for each quark (antiquark), and $s_i$ and $s_j$ represent the spin operators for the $i^{th}$ and $j^{th}$ quarks (antiquark) and $b_{ij}$ is specified as:
 \begin{equation}
     b_{ij} = \frac{16\pi\alpha_s}{9 m_i m_j} \bra{\Psi_0}\delta^3(\Vec{r})\ket{\Psi_0}
 \end{equation}
 where $\Psi_0$ denotes the hidden-bottom pentaquark ground state wavefunction. Therefore, effective mass equations for several quarks inside the pentaquark are:
 \begin{equation}
 m_1^{eff} = m_1 + \alpha b_{12} + \beta b_{13} + \gamma b_{14} + \eta b_{15}
 \end{equation}
 
 \begin{equation}
  m_2^{eff} = m_2 + \alpha b_{12} + \beta^{'} b_{23} + \gamma^{'} b_{24} + \eta^{'} b_{25}
  \end{equation}

   \begin{equation}
   m_3^{eff} = m_3 + \beta b_{13} + \beta^{'} b_{23} + \gamma^{''} b_{34} + \eta^{''} b_{35} 
   \end{equation}

   \begin{equation}
    m_4^{eff} = m_4 + \gamma b_{14} + \gamma^{'} b_{24} + \gamma^{''} b_{34} + \eta^{'''} b_{45} 
 \end{equation}

  \begin{equation}
    m_5^{eff} = m_5 + \eta b_{15} + \eta^{'} b_{24} + \eta^{''} b_{34} + \eta^{'''} b_{45} 
 \end{equation}
 Here, the numbers 1, 2, 3, 4, and 5 represent the $u$, $d$, $s$, $c$, and $b$ quarks. If we include two/three/four/five identical quarks, these equations will be altered. We can define the product of spin quantum numbers as \cite{Rohit}:
\begin{align*}
      s_{i}\cdot s_{j}=&  +1/4 \rightarrow \hspace{0.3cm}\uparrow \uparrow\\
       =& -1/2 \rightarrow \hspace{0.3cm} \uparrow \downarrow\\
       =& -1/4 \rightarrow \hspace{0.3cm}\downarrow \downarrow
\end{align*}
In Ref. \cite{sharma2024}, spin-spin interactions and parameters are thoroughly examined using Eqs. \eqref{10} and \eqref{11}. By substituting the values of spin-spin interactions and associated parameters, a modified form of the effective mass equation for $J^P = 5/2^-$ pentaquarks with spin wavefunction ($\uparrow\uparrow\uparrow\uparrow\uparrow$) described as:
\begin{align}
M_{P_{{5/2}^-}} = m_1 + m_2 + m_3 + m_4 + m_5 + \frac{b_{12}}{4}  + \frac{b_{13}}{4}  + \frac{b_{14}}{4} \nonumber \\
 + \frac{b_{15}}{4}  + \frac{b_{23}}{4}  + \frac{b_{24}}{4}  + \frac{b_{25}}{4}  + \frac{b_{34}}{4}  + \frac{b_{35}}{4}  + \frac{b_{45}}{4}
 \label{52}
\end{align}
Similarly, the total mass of a pentaquark for $J^P = 3/2^-$ with spin wavefunction ($\uparrow\uparrow\uparrow\uparrow\downarrow$) is defined as follows:
\begin{align}
M_{P_{{3/2}^-}} = m_1 + m_2 + m_3 + m_4 + m_5 + \frac{b_{12}}{4}  + \frac{b_{13}}{4}  + \frac{b_{14}}{4} \nonumber \\
 - \frac{b_{15}}{2}  + \frac{b_{23}}{4}  + \frac{b_{24}}{4}  - \frac{b_{25}}{2}  + \frac{b_{34}}{4}  - \frac{b_{35}}{2}  - \frac{b_{45}}{2}
 \label{32}
\end{align}
 
The quark masses are obtained from Ref. \cite{Rohit}, and the hyperfine interaction terms $b_{ij}$ are computed using the equations for total effective masses \ref{52} and \ref{32}. They are expressed as follows:

\begin{align}
     m_u = m_d = \hspace{0.3cm} 362 MeV, \hspace{0.3cm} m_s = \hspace{0.3cm} 539 MeV \nonumber \\
    m_c = \hspace{0.3cm} 1710 MeV, \hspace{0.3cm} m_b = \hspace{0.3cm} 5043 MeV
\end{align}

\begin{equation}
     b_{uu} = \hspace{0.3cm} b_{ud} = \hspace{0.3cm} b_{dd} =165.9 MeV
 \end{equation}

 \begin{equation}
     b_{us} = \hspace{0.3cm} b_{ds} \hspace{0.3cm} = \left(\frac{m_u}{m_s}\right)b_{uu} = 111.13 MeV
 \end{equation}

\begin{equation}
     b_{ss} = \hspace{0.3cm} \left(\frac{m_u}{m_s}\right)^2 b_{uu} =  74.83 MeV
 \end{equation}

\begin{equation}
     b_{sb} = \hspace{0.3cm} \left(\frac{m_u^2}{m_s m_b}\right) b_{uu} = 7.99 MeV
\end{equation}

\begin{equation}
     b_{bb} = \hspace{0.3cm} \left(\frac{m_u}{ m_b}\right)^2 b_{uu} = 0.85 MeV
 \end{equation}

\begin{equation}
     b_{ub} = \hspace{0.3cm} b_{db} = \hspace{0.3cm} \left(\frac{m_u}{m_b}\right)b_{uu}  = 11.9 MeV
 \end{equation}

 The effective quark masses for hidden-bottom pentaquarks with $J^P$ = $3/2^-$ and $5/2^-$ can be determined using the hyperfine interaction terms $b_{ij}$ and quark masses as follows: \\
 \\
i) For ($s$ = 0) octet pentaquarks,
\begin{equation}
    m_u^* = \hspace{0.3cm} m_d^* = 401.98 MeV
    \end{equation}
    \begin{equation}
    m_b^* = 5047.25, \hspace{0.3cm} m_{\Bar{b}}^* = 5033.86 MeV
    \end{equation}
\\
(ii) For ($s$ = 1) octet pentaquarks,
\begin{equation}
    m_u^* = \hspace{0.3cm} m_d^* = 395.17 MeV
    \end{equation}
    \begin{equation}
        m_s^* = \hspace{0.3cm} 565.85 MeV
    \end{equation}
    \begin{equation}
    m_b^* = 5046.76, \hspace{0.3cm} m_{\Bar{b}}^* = 5034.84 MeV
    \end{equation}
\\

(iii) For ($s$ = 2) octet pentaquarks,
\begin{equation}
    m_u^* = \hspace{0.3cm} m_d^* = 388.36 MeV
    \end{equation}
    \begin{equation}
        m_s^* = \hspace{0.3cm} 561.28 MeV
    \end{equation}
    \begin{equation}
    m_b^* = 5046.27, \hspace{0.3cm} m_{\Bar{b}}^* = 5035.82 MeV
    \end{equation}
\\
Similarly, for decuplet particles having $J^P = 5/2^-$,
 \\
i) For ($s$ = 0) decuplet pentaquarks,
\begin{equation}
    m_u^* = \hspace{0.3cm} m_d^* = 406.45 MeV
    \end{equation}
    \begin{equation}
    m_b^* = \hspace{0.3cm} m_{\Bar{b}}^* = 5047.57 MeV
    \end{equation}
\\
(ii) For ($s$ = 1) decuplet pentaquarks,
\begin{equation}
    m_u^* = \hspace{0.3cm} m_d^* = 399.64 MeV
    \end{equation}
    \begin{equation}
        m_s^* = \hspace{0.3cm} 568.84 MeV
    \end{equation}
    \begin{equation}
    m_b^* = \hspace{0.3cm} m_{\Bar{b}}^* = 5047.08 MeV
    \end{equation}
\\
(iii) For ($s$ = 2) decuplet pentaquarks,
\begin{equation}
    m_u^* = \hspace{0.3cm} m_d^* = 392.82 MeV
    \end{equation}
    \begin{equation}
        m_s^* = \hspace{0.3cm} 564.27 MeV
    \end{equation}
    \begin{equation}
    m_b^* = \hspace{0.3cm} m_{\Bar{b}}^* = 5046.59 MeV
    \end{equation}
\\
(iv) For ($s = 3$) decuplet baryons,
    \begin{equation}
        m_s^* = \hspace{0.3cm} 559.70 MeV
    \end{equation}
    \begin{equation}
    m_b^* = \hspace{0.3cm} m_{\Bar{b}}^* = 5046.10 MeV
    \end{equation}
 The hidden-bottom pentaquark masses for both octet and decuplet configurations using the effective mass scheme are reported in Table \ref{tab:3} and \ref{tab:4} respectively. In the following section, we presented the screened charge scheme for computing the magnetic moments of hidden-bottom pentaquarks.
\subsection{Screened Charge Scheme}
The charge of a quark within an exotic hadron may undergo modification analogous to the alterations observed in its mass due to the surrounding environment. This phenomenon arises from the interplay between the quark's intrinsic properties and the strong force interactions with neighboring particles. In certain contexts, such as within a dense nuclear medium or in the presence of external fields, the charge of a quark can become effectively screened or modified, leading to interesting phenomena. Understanding these dynamics is crucial for decoding the behavior of exotic hadrons and explaining their role in the complex aspect of particle physics. Hence, the effective charge of quark 'a' in the pentaquark (a, f, x, y, z) can be expressed as: 
\begin{equation}
    e_a^P = e_a + \alpha_{af} e_f + \alpha_{ax} e_x + \alpha_{ay} e_y + \alpha_{az} e_z
    \label{Screened Charge1}
    \end{equation}
Similarly, effective charge equations for other quarks (antiquarks) are defined as:
\begin{equation}
    e_f^P = e_f + \alpha_{fa} e_a + \alpha_{fx} e_x + \alpha_{fy} e_y + \alpha_{fz} e_z,
    \end{equation}

    \begin{equation}
    e_x^P = e_x + \alpha_{xa} e_a + \alpha_{xf} e_f + \alpha_{xy} e_y + \alpha_{xz} e_z,
    \end{equation}

\begin{equation}
    e_y^P = e_y + \alpha_{ya} e_a + \alpha_{yf} e_f + \alpha_{yx} e_x + \alpha_{yz} e_z,
    \end{equation}

    \begin{equation}
    e_z^P = e_z + \alpha_{za} e_a + \alpha_{zf} e_f + \alpha_{zx} e_x + \alpha_{zy} e_y
    \label{screened5}
    \end{equation}
where the quark charges are denoted by $e_a$, $e_f$, $e_x$, $e_y$, and $e_z$. When we take the isospin symmetry into account, we obtain several relations:
\begin{align}
    \alpha_{af} = \alpha_{fa}, \hspace{0.3cm}\alpha_{ax} = \alpha_{xa}, \hspace{0.3cm} \alpha_{ay} = \alpha_{ya}, \hspace{0.3cm} \alpha_{az} = \alpha_{za}
\end{align}
Thus,
\begin{align}
    \alpha_{uu} = \alpha_{ud} = \alpha_{dd} = \alpha_1 \\ \nonumber
    \alpha_{us} = \alpha_{ds} = \beta_1 \\ \nonumber
    \alpha_{ss} = \beta_2
\end{align}
For the bottom sector,
 \begin{align}
     \alpha_{ub} = \alpha_{db} = \beta_3, \hspace{0.3cm} \alpha_{sb} = \alpha_2 \\ \nonumber
     \alpha_{cb} = \beta_4, \hspace{0.3cm} \alpha_{bb} = \alpha_3
 \end{align}
Further, the reduction of these parameters can be achieved by applying SU(3) symmetry.
\begin{equation}
    \alpha_1 = \beta_1 = \beta_2
\end{equation}
Using the Ansatz formalism, we can calculate the screening parameter $\alpha_{ij}$ as: 
\begin{equation}
    \alpha_{ij} = \mid{\frac{m_i - m_j}{m_i + m_j}}\mid \times \delta
\end{equation}
where $\delta$ = 0.81 \cite{Bains} and $m_i$ and $m_j$ are the corresponding quark masses ($i, j = u, d, s, c, b$). The magnetic moments of hidden-bottom pentaquark states can be predicted using the values of these parameters. By placing these parameters in effective charge equations \eqref{Screened Charge1}-\eqref{screened5}, and by presenting the magnetic moment operator as:
\begin{equation}
    \mu = \sum_i \frac{e_i^P}{2 m_i^{eff}} \sigma_i
\end{equation}
we can calculate the magnetic moments of pentaquarks. The magnetic moment of the multiquark system is divided into two parts:
 \begin{equation}
     \Vec{\mu} = \Vec{\mu}_{spin} + \Vec{\mu}_{orbit}
 \end{equation}
which can be written as \cite{sharma2024}:
\begin{equation}
    \Vec{\mu} = \hspace{0.3cm} \sum_i \mu_i (2 \Vec{s_i} + \Vec{l_i}) = \hspace{0.3cm} \sum_i \mu_i(2\Vec{s_i}) = \hspace{0.3cm} \sum_i \mu_i(\Vec{\sigma_i})
\label{Magnetic moment}
\end{equation}
As orbital excitation is absent, the magnetic moment is only determined by the spin component. Magnetic moments can be determined by calculating the expectation value of Eq.\eqref{Magnetic moment} using the pentaquark spin-flavor wavefunction as:
\begin{equation}
    \mu = \bra{\Psi_{sf}}\Vec{\mu}\ket{\Psi_{sf}}
    \label{magnetic123}
\end{equation}
Now, we define the spin and flavor wavefunctions of the hidden-bottom pentaquarks having $J^P = 3/2^-$ and $5/2^-$ respectively. For spin- $3/2
^-$ case, there are a total of 4 spin symmetries available. We used the $\chi_2$ spin symmetry because it provides less fluctuations in magnetic moment values for both the schemes introduced. Therefore, the spin-flavor wavefunction for spin-$3/2^-$ hidden-bottom pentaquarks using the group theory approach is developed through the utilization of the $SU_{sf}(6)$ spin-flavor representation as $SU_{sf}(6)$ = $SU_{f}(3) \otimes SU_{s}(2)$ and can be written as:
 \begin{equation}
     \Psi_{sf} =  \chi_2 \otimes \phi_f 
 \end{equation}
where the definition of the spin wavefunction for spin-3/2 is \cite{wavefunction}:
\begin{center}
    $\chi_2 = \frac{1} {2\sqrt{5}}\ket{4\uparrow\uparrow\uparrow\uparrow\downarrow - (\uparrow\uparrow\uparrow\downarrow + \uparrow\uparrow\downarrow\uparrow + \uparrow\downarrow\uparrow\uparrow + \downarrow\uparrow\uparrow\uparrow)\uparrow}$
\end{center}
 The flavor wavefunction for the state $\ket{uudb\Bar{b}}$ is defined as:
\begin{equation}
 \phi_f = \frac{-1}{\sqrt{6}}|dbuu +ubdu + ubud - bduu - budu - buud\rangle\Bar{b}
\end{equation} 
by substituting this in magnetic moment expression, we get for the case of $uudb\Bar{b}$:
\begin{equation}
    \mu = \frac{9}{10}[2\mu_u + \mu_d + \mu_b]- \frac{3}{5}\mu_{\Bar{b}}
\end{equation}
We can also establish this for other octet particles. Now, for the decuplet particles having spin-$5/2^-$, \begin{equation}
     \Psi_{sf} =  \chi_1 \otimes \phi_f 
 \end{equation}

For spin-5/2, the spin wavefunction is defined as:
\begin{center}
    $\chi_1 = \ket{\uparrow\uparrow\uparrow\uparrow\uparrow}$
\end{center}
The flavor wavefunction for $\ket{uuub\Bar{b}}$ is defined as follows:

\begin{align}
 \phi_f = \frac{1}{2\sqrt{5}}|uuub\Bar{b} +\Bar{b}uubu + u\Bar{b}ubu + uu\Bar{b}bu + uuu\Bar{b}b + uubu\Bar{b}\nonumber \\ + \Bar{b}ubuu + u\Bar{b}buu + uu\Bar{b}ub + uub\Bar{b}u + ubuu\Bar{b} + \Bar{b}buuu + u\Bar{b}uub \nonumber \\ + ub\Bar{b}uu + ubu\Bar{b}u + buuu\Bar{b} + \Bar{b}uuub + b\Bar{b}uuu + bu\Bar{b}uu + buu\Bar{b}u\rangle
\end{align}
By substituting this spin-flavor wavefunction in magnetic moment expression, we get:
\begin{equation}
    \mu =  \mu_u + \mu_u + \mu_u + \mu_b + \mu_{\Bar{b}}
\end{equation}
Where $\mu_u$, $\mu_b$, and $\mu_{\Bar{b}}$ are the magnetic moments for quarks $u$, $b$, and $\Bar{b}$ respectively. Likewise, we may establish this for various decuplet particles. We determined the magnetic moments of the hidden-bottom octet and decuplet of pentaquarks by utilizing the spin-flavor wavefunction in equation \eqref{magnetic123}. Using the relations described above, one can follow the expressions for the magnetic moments of pentaquarks, which are written in Table.  \ref{tab: expression} and \ref{tab: expressions} for octet and decuplet particles respectively. In the next section, we briefly discussed the two possible multiplets of hidden-bottom pentaquarks i.e. octet and decuplet respectively.

\begin{table*}[ht!]
        \centering
         \caption{Table displaying the masses of hidden-bottom Pentaquarks in the $SU(3)_f$ octet configuration using the extension of Gursey-Radicati mass formula and the effective mass scheme.The nomenclature for pentaquark states is identical to that seen in Figure \ref{fig:1}. All masses are expressed in MeV units.}
         \tabcolsep 1.0mm 
          \begin{tabular}{cccccc}
      \hline
      \hline
   Pentaquark States & G-R Formula &  Effective Mass   &  Ref. \cite{magneticmom} &  Ref. \cite{octetcomp} & Ref. \cite{chiral}\\
     \hline
     $P_b^{00}$, $P_b^{0+}$ &  11308.10 $\pm$ 69.08  & 11287.1  & 11412  & 11130 & 11124 \\ 
     \hline
     $P_b^{1-}$, $P_b^{10}$, $P_b^{1+}$  & 11514.40 $\pm$ 69.12  & 11437.8 & 11510 &  11255 & - \\ 
     \hline
     $P_b^{20}$, $P_b^{2-}$ &  11624.70 $\pm$ 69.08 & 11593.0  & 11651 &  11365 & - \\ 
     \hline
     $P_b^{1^{'}0}$ &  11450.40 $\pm$ 69.07  & 11437.8 & 11490 &   11240 & - \\ 
     \hline
     \hline
       \end{tabular}
    \label{tab:3}
\end{table*}

\begin{table}[]
    \centering
     \caption{Magnetic moment expressions for hidden-bottom pentaquarks ($J^P = 3/2^-$) expressed in terms of  effective mass scheme (in $\mu_N$)
.}
    \begin{tabular}{cc}
    \toprule
    \toprule
       Octet States  & \hspace{0.3cm} Effective mass Scheme  \\
       \midrule
        $P_b^{0+}$  & \hspace{0.3cm}  $\mu = \frac{9}{10}[2\mu_u + \mu_d + \mu_b]- \frac{3}{5}\mu_{\Bar{b}}$\\
         \midrule
        $P_b^{00}$ & \hspace{0.3cm}  $\mu = \frac{9}{10}[\mu_u + 2\mu_d + \mu_b]- \frac{3}{5}\mu_{\Bar{b}}$ \\
         \midrule
        $P_b^{1+}$ & \hspace{0.3cm}  $\mu = \frac{9}{10}[2\mu_u + \mu_s + \mu_b]- \frac{3}{5}\mu_{\Bar{b}}$ \\
      
         \midrule
        $P_b^{1-}$ & \hspace{0.3cm} $\mu = \frac{9}{10}[2\mu_d + \mu_s + \mu_b]- \frac{3}{5}\mu_{\Bar{b}}$  \\
         \midrule
        $P_b^{10}$, $P_b^{1^{'}0}$ & \hspace{0.3cm}  $\mu = \frac{9}{10}[\mu_u + \mu_d + \mu_s + \mu_b]- \frac{3}{5}\mu_{\Bar{b}}$  \\
         \midrule
        $P_b^{20}$ & \hspace{0.3cm}  $\mu = \frac{9}{10}[\mu_u + 2\mu_s + \mu_b]- \frac{3}{5}\mu_{\Bar{b}}$  \\
        \midrule
        $P_b^{2-}$ & \hspace{0.3cm}  $\mu = \frac{9}{10}[\mu_d + 2\mu_s + \mu_b]- \frac{3}{5}\mu_{\Bar{b}}$ \\
        \bottomrule
        \bottomrule
    \end{tabular}
  \label{tab: expression}
\end{table}

\begin{table*}
    \centering
    \caption{Magnetic moments assignments of hidden-bottom octet of pentaquarks with $J^P = 3/2^-$ utilising the effective mass scheme, screened charge scheme, and combining the effective mass and screened charge schemes.}
    \tabcolsep 1.0mm 
    \begin{tabular}{p{2cm}ccccc}
    \hline
    \hline
     State &  Quark Content &  Effective mass  &  Screened Charge &  Eff. mass + Screen. Charge & Ref. \cite{magneticmom} \\
     \hline  
$P_b^{+(0)}$ &  $uudb\Bar{b}$ & 2.00 & 2.28  & 2.04 & 2.78 \\
       \hline
$P_b^{0(0)}$ &  $uddb\Bar{b}$ &  -0.09 &  -0.09 &  -0.09 & -0.04\\
         \hline
 $P_b^{+(1)}$ &  $uusb\Bar{b}$ & 2.26  & 2.62  & 2.39  &  2.60\\
          \hline
 $P_b^{0(1)}$, $P_b^{1^{'}0}$ &  $udsb\Bar{b}$ & 0.12 & 0  & -0.02 & 0.10\\
           \hline
 $P_b^{-(1)}$ &  $ddsb\Bar{b}$& -2.01  & -2.62  & -2.43  & -2.41\\
            \hline
$P_b^{0(2)}$ &  $ussb\Bar{b}$ & 0.36  & 0.50  & 0.45 &  0.51\\
             \hline
$P_b^{-(2)}$ & $dssb\Bar{b}$& -1.81 & -2.36  & -2.24 & -2.43\\
\hline
\hline
    \end{tabular}
    \label{tab: magnetic moments of octet}
\end{table*}

\section{Result and Discussion}
\subsection{Hidden-Bottom Octet}
\begin{figure}[h]
    \centering
     \caption{Hidden-bottom octet of pentaquarks. $Y$ and $I_3$ represent the hypercharge and third component of isospin respectively. Each state is designated as $P_b^{sq}(m)$, where $s$ represents the strangeness, $q$ represents the charge and $m$ represents the mass of the hidden-bottom pentaquarks.}
    \includegraphics[width=9cm]{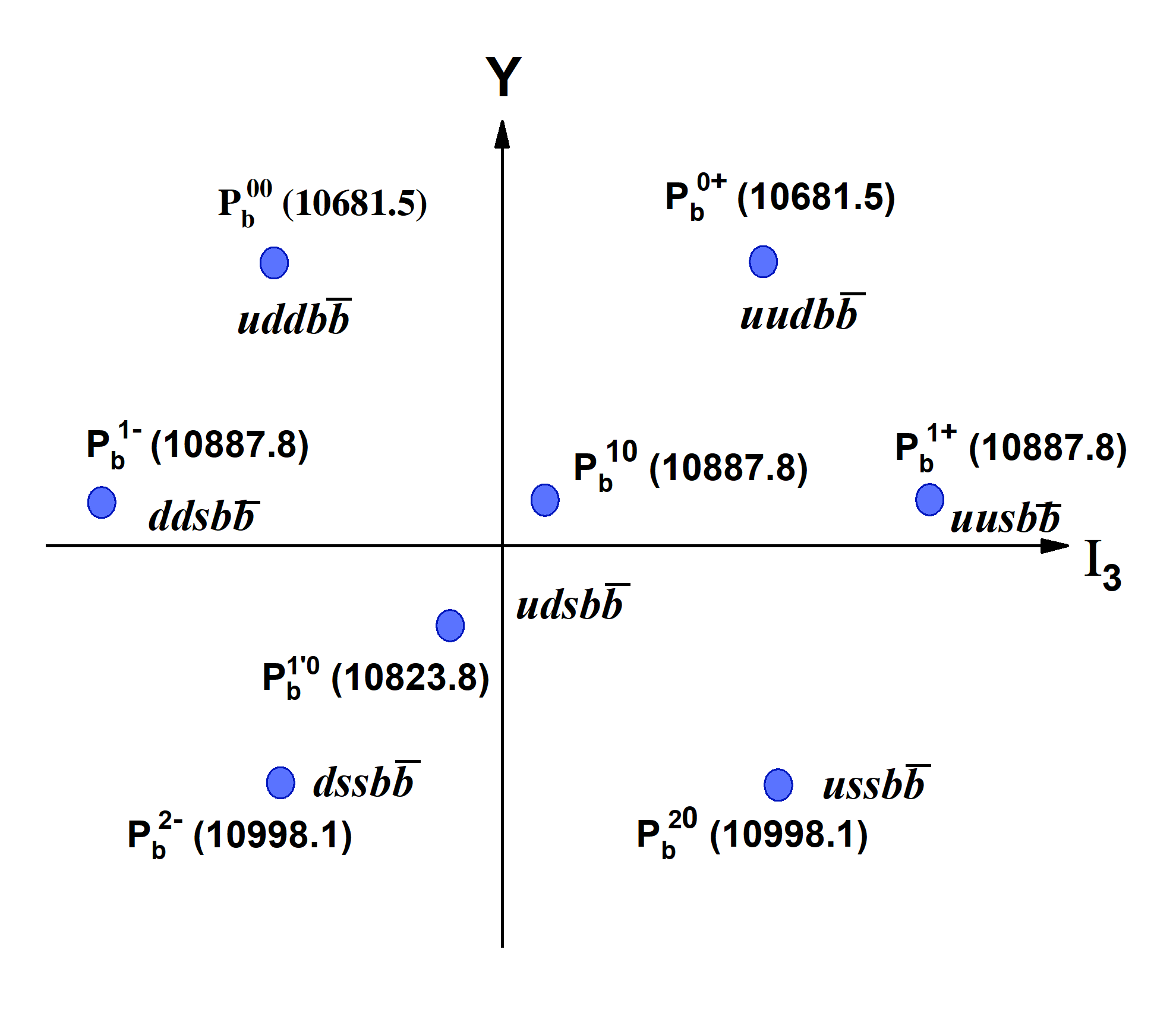}
    \label{fig:1}
\end{figure}

We classified the hidden-bottom pentaquarks into octet form by utilizing the SU(3) flavor representation. The SU(3) flavor symmetry makes specific predictions about the masses and characteristics of exotic baryons in the octet. A fundamental prediction is that pentaquarks containing the same SU(3) flavor components should exhibit similar masses. The SU(3) flavor symmetry plays a critical role in comprehending the arrangement and characteristics of pentaquarks. It offers a systematic approach to classify these particles according to their quark composition. The octet representation for hidden-bottom pentaquarks is shown in Figure \ref{fig:1} with their respective notations and quark contents. The assigned $J^P$ value for the hidden-bottom octet is $3/2^-$, which is similar to the work proposed in Ref. \cite{Santo} for hidden-charm configuration. The estimated masses of hidden-bottom octet particles are calculated using the extended Gursey-Radicati mass formula and effective mass scheme, as shown in the table \ref{tab:3}. As shown in Figure \ref{fig:1}, $P_b^{00}$ and $P_b^{0+}$ are the isospin doublet states and thus have the same mass. Similarly, $P_b^{1+}$, $P_b^{10}$, and $P_b^{1-}$ are the isospin partners, resulting in identical mass. Further, $P_b^{20}$ and $P_b^{2-}$ form isospin doublet, and $P_b^{1^{'}0}$ is an isospin singlet state. The magnetic moment expressions for the hidden-bottom octet of pentaquarks are listed in the table. \ref{tab: expression} and magnetic moment assignments of the octet form using the effective mass and screened charge schemes are reported in Table \ref{tab: magnetic moments of octet}. We compared our estimated masses with Ref. \cite{octetcomp}, \cite{chiral}, and \cite{magneticmom} and they are in reasonable agreement. Our examination of the magnetic moments is also consistent with Ref. \cite{magneticmom} and will be beneficial for forthcoming experimental investigations. 

\subsection{Hidden-Bottom Decuplet}
\begin{figure}[ht!]
    \centering
     \caption{Hidden-bottom decuplet of pentaquarks. $Y$ and $I_3$ represent the hypercharge and third component of isospin respectively. We named each state as $P_b^{sq}(m)$, where $s$ represents the strangeness, $q$ represents charge and $m$ represents mass of the pentaquarks.}
    \includegraphics[width=9cm]{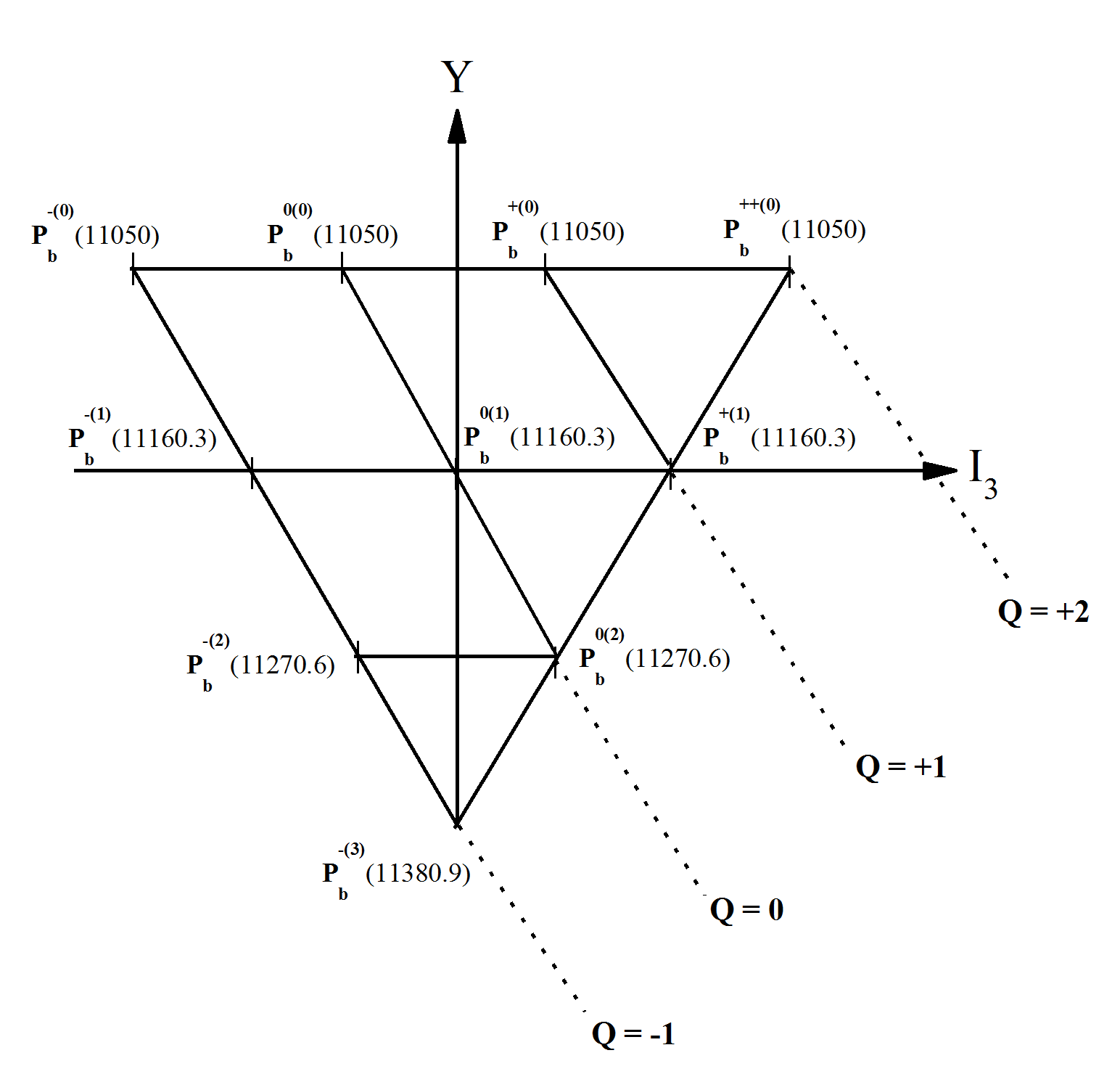}
    \label{fig:2}
\end{figure}

This section discusses the hidden-bottom decuplet configuration of the SU(3) representation. The $J^P$ value assigned for the ground state decuplet particles is $5/2^-$. The decuplet representation is shown in Figure \ref{fig:2} and masses of the decuplet particles using the extended version of the GR mass formula and the eff. mass scheme are reported in Table \ref{tab:4}. As shown in Figure \ref{fig:2}, $P_b^{++(0)}$, $P_b^{+(0)}$,  $P_b^{0(0)}$, and $P_b^{-(0)}$ forms isospin quartet and have same mass. In a similar manner,  $P_b^{+(1)}$, $P_b^{0(1)}$, and $P_b^{-(1)}$ forms isospin triplet and thus have identical mass. Further,  $P_b^{0(2)}$ and $P_b^{-(2)}$ are isospin doublet particles and $P_b^{-(3)}$ are isospin singlet state. Also, the study of magnetic moments of hidden-bottom decuplet of particles is carried out using the effective mass and screened charge schemes. The corresponding expressions for magnetic moments are written in Table \ref{tab: expressions} and their magnetic moments are reported in Table \ref{tab: magnetic moments}. We compared our predicted masses with Ref. \cite{colormagnetic}, \cite{chiral}, and \cite{magneticmom}, and they are in reasonable agreement. Further, the comparison of magnetic moments with Ref. \cite{magneticmom} also shows excellent agreement and will be beneficial for future experimental research. In the next section, we studied the partial widths of the hidden-bottom octet of pentaquark with $J^P = 3/2^-$ by adopting the effective Lagrangian method.

\begin{table*}[ht!]
        \centering
         \caption{Table for masses of predicted hidden-bottom Pentaquark Decuplet using the extended form of the Gursey-Radicati mass formula and the effective mass scheme. The nomenclature for pentaquark states is identical to that seen in the figure.\ref{fig:2}. All masses are measured in MeV.}
         \tabcolsep 0.8mm 
          \begin{tabular}{cccccc}
      \hline
      \hline
   Pentaquark States &  G-R Formula &  Eff. Mass Scheme & Ref. \cite{colormagnetic} &  Ref. \cite{chiral} & Ref. \cite{magneticmom}\\
     \hline
     $P_b^{++(0)}$, $P_b^{+(0)}$,  $P_b^{0(0)}$, $P_b^{-(0)}$ &   
 11676.60 $\pm$ 70.20   & 11314.1  &   11246.5 &  11052 & 11235 \\ 
 \hline
     $P_b^{+(1)}$, $P_b^{0(1)}$, $P_b^{-(1)}$  & 11786.90 $\pm$ 70.09  & 11462.2  & 11429.9 & -  & 11524  \\
     \hline
     $P_b^{0(2)}$, $P_b^{-(2)}$ &   11897.20 $\pm$ 70.06 & 11614.4   & 11575.7 & 11141 & 11669\\ 
     \hline
     $P_b^{-(3)}$ & 12007.50 $\pm$ 70.10  & 11771.2 & - & - & 11673 \\ 
     \hline
     \hline
       \end{tabular}
    \label{tab:4}
\end{table*}

\begin{table}[]
    \centering
     \caption{Magnetic moment expressions for hidden-bottom pentaquarks with a ($J^P = 5/2^-$) states by utilising effective mass scheme (in $\mu_N$).}
    \begin{tabular}{cc}
    \hline
    \hline
       Decuplet States  & \hspace{0.3cm} Effective mass Scheme  \\
       \hline
        $P_b^{++(0)}$  & \hspace{0.3cm} 3 $\mu_u^{eff}$ + $\mu_b^{eff}$ + $\mu_{\Bar{b}}^{eff}$\\
\hline
        $P_b^{+(0)}$ & \hspace{0.3cm} 2$\mu_u^{eff}$ + $\mu_d^{eff}$ + $\mu_b^{eff}$ + $\mu_{\Bar{b}}^{eff}$ \\
        \hline
        $P_b^{0(0)}$ & \hspace{0.3cm} $\mu_u^{eff}$ + 2$\mu_d^{eff}$ + $\mu_b^{eff}$ + $\mu_{\Bar{b}}^{eff}$ \\
        \hline
        $P_b^{-(0)}$ & \hspace{0.3cm} 3$\mu_d^{eff}$ + $\mu_b^{eff}$ + $\mu_{\Bar{b}}^{eff}$  \\
        \hline
        $P_b^{+(1)}$ & \hspace{0.3cm} 2$\mu_u^{eff}$ + $\mu_s^{eff}$ + $\mu_b^{eff}$ + $\mu_{\Bar{b}}^{eff}$  \\
        \hline
        $P_b^{0(1)}$ & \hspace{0.3cm} $\mu_u^{eff}$ + $\mu_d^{eff}$ + $\mu_s^{eff}$ + $\mu_b^{eff}$ + $\mu_{\Bar{b}}^{eff}$ \\
        \hline
        $P_b^{-(1)}$ & \hspace{0.3cm} 2$\mu_d^{eff}$ + $\mu_s^{eff}$ + $\mu_b^{eff}$ + $\mu_{\Bar{b}}^{eff}$   \\
        \hline
        $P_b^{0(2)}$ & \hspace{0.3cm}  $\mu_u^{eff}$ + 2$\mu_s^{eff}$ + $\mu_b^{eff}$ + $\mu_{\Bar{b}}^{eff}$  \\
        \hline
        $P_b^{-(2)}$ & \hspace{0.3cm} $\mu_d^{eff}$ + 2$\mu_s^{eff}$ + $\mu_b^{eff}$ + $\mu_{\Bar{b}}^{eff}$  \\
        \hline
        $P_b^{-(3)}$ & \hspace{0.3cm}  3$\mu_s^{eff}$ + $\mu_b^{eff}$ + $\mu_{\Bar{b}}^{eff}$ \\
        \hline
        \hline
    \end{tabular}
  \label{tab: expressions}
\end{table}

\begin{table*}
    \centering
    \caption{Investigating the magnetic moments of the hidden-bottom decuplet of pentaquarks with $J^P = 5/2^-$ utilizing the effective mass scheme, screened charge scheme, and a combination of effective mass and screened charge schemes.}
    \tabcolsep 0.8mm 
    \begin{tabular}{p{1cm}ccccc}
    \hline
    \hline
     States &  Quark Contents &  Eff. mass  &  Screen. Charge &  Eff. mass + Screen. Charge & Ref. \cite{magneticmom} \\
     \hline
$P_b^{++(0)}$ &  $uuub\Bar{b}$ &  4.62 & 5.70 &  5.13 & -\\
   \hline   
$P_b^{+(0)}$ &  $uudb\Bar{b}$ & 2.31 &  2.85 &  2.56 & - \\
       \hline
$P_b^{0(0)}$ &  $uddb\Bar{b}$ &  0 &  0 &  0 & -\\
         \hline
 $P_b^{-(0)}$ &  $dddb\Bar{b}$ &  -2.31 &  -2.85 &  -2.56 & -\\
         \hline
 $P_b^{+(1)}$ &  $uusb\Bar{b}$ &  2.58 &  3.23 &  2.94 & 3.19 \\
          \hline
 $P_b^{0(1)}$ &  $udsb\Bar{b}$ &  0.23 &  0.10 &  0.07 & 0.24\\
           \hline
 $P_b^{-(1)}$ &  $ddsb\Bar{b}$&  -2.11 &  -3.02 &  -2.79 & -2.70\\
            \hline
$P_b^{0(2)}$ &  $ussb\Bar{b}$ &  0.48 &  0.67 &  0.59 &  0.49\\
             \hline
$P_b^{-(2)}$ & $dssb\Bar{b}$& -1.90 &  -2.73  & -2.58 & -2.46\\
              \hline
$P_b^{-(3)}$ & $sssb\Bar{b}$&  -1.67 &  -1.98 & -1.92 & -\\
\hline
\hline
    \end{tabular}
    \label{tab: magnetic moments}
\end{table*}

\section{Partial Decay Widths}
In this section, we determined the partial widths for the SU(3) flavor octet configuration of hidden-bottom pentaquarks. The study of decay widths contributes to a better understanding of resonances, their internal structure, and interactions. The study of decay width also provides information about the lifetime of a particle. To compute the partial widths, we applied the effective Lagrangian approach for $P_bN\Upsilon$ couplings from Ref. \cite{DECAY} as follows:\\
 \begin{align}
L_{PN\Upsilon}^{3/2^-} &= i \Bar{P_\mu}\left[\frac{g_1}{(2M_N)}N\Gamma_\nu{^-}\right] \Upsilon^{\mu\nu}-i\Bar{P_\mu} \bigg[\frac{ig_2}{(2M_N)^2}\Gamma_\nu^{-} \partial_\nu N + \frac{ig_3}{(2M_N)^2}\Gamma_\nu^{-} N \partial_\nu \bigg]\Upsilon^{\mu\nu} + H.C.,
\label{Lagrangian}
 \end{align}
 where $P$ is the field for pentaquarks with $J^P = 3/2^-$, $N$ and $\Upsilon$ are the nucleon and the bottomonium fields, respectively. The $\Gamma$ matrices are defined as follows:\\
 \begin{equation}
     \Gamma_\nu^- = \begin{pmatrix}
      \gamma_\nu \gamma_5 \\
  \gamma_\nu \\
  
\end{pmatrix}, \hspace{0.5cm} \Gamma^+ = \begin{pmatrix}
  \gamma_5 \\
  1 \\
 \end{pmatrix}
 \end{equation}
Wang $et$  $al$ \cite{Wang} observed that the final momenta of the pentaquark state decaying into $J/\psi N$ are significantly smaller than the nucleon mass. Therefore, in the case of hidden-bottom pentaquarks decaying into $\Upsilon N$, Final state momenta are significantly reduced because of the substantial mass of the decay products, allowing greater partial wave components corresponding to $(p/M_N)^2$ to be disregarded, so only first term in Eq. \eqref{Lagrangian} is being taken into account. This approximation results in decay of $P_b^{+(0)}$ in the $N \Upsilon$ channel \cite{oh2011role}:
 \begin{align}
     \Gamma(P_b^{+(0)} \rightarrow N \Upsilon) = \frac{\overline{g}_{{N}\Upsilon}^2}{12\pi}\frac{p_{N}}{M_{P_b^{+(0)}}}(E_{N} + M_{N}) 
[2(E_{N})(E_{N} - M_{N}) + (M_{P_b^{+(0)}} - M_{N})^2 + 2(M_{\Upsilon})^2]
 \end{align}
 with coupling constant $\overline{g}_{{N}\Upsilon}$ defined as:
 \begin{equation}
\overline{g}_{{N}\Upsilon} = \frac{g_1}{(2M_{N})}
 \end{equation}
 where, energy and momentum $E_{N}$ and $p_{N}$ are  $E_{N} = (M_{P_b^{+(0)}}^2+M_{N}^2-M_{\Upsilon}^2)/(2M_{P_b^{+(0)}})$ and $p_{N} = \sqrt{E_{N}^2-M_{N}^2}$ are the kinematic parameters. \\
 Expressions for partial decay widths of pentaquarks with strangeness are shown as:
\begin{align}
     \Gamma(P_b^{+(1)} \rightarrow \Sigma^{+} \Upsilon) = \frac{\overline{g}_{{\Sigma^{+}}\Upsilon}^2}{12\pi}\frac{p_{\Sigma^{+}}}{M_{P_b^{+(1)}}}(E_{\Sigma^{+}} + M_{\Sigma^{+}}) 
[2(E_{\Sigma^{+}})(E_{\Sigma^{+}} - M_{\Sigma^{+}}) + (M_{P_b^{+(1)}} - M_{\Sigma^{+}})^2 + 2(M_{\Upsilon})^2]
 \end{align}
 with coupling constant $\overline{g}_{{\Sigma^{+}}\Upsilon}$:
 \begin{equation}
\overline{g}_{{\Sigma^{+}}\Upsilon} = \frac{g_1}{(2M_{\Sigma^{+}})}
 \end{equation}
 Similarly, the expression for doubly strange states ($P_b^{0(2)}$) for partial width are written as:
 \begin{align}
     \Gamma(P_b^{0(2)} \rightarrow \Xi^{0} \Upsilon) = \frac{\overline{g}_{{\Xi^{0}}\Upsilon}^2}{12\pi}\frac{p_{\Xi^{0}}}{M_{P_b^{0(2)}}}(E_{\Xi^{0}} + M_{\Xi^{0}})
[2(E_{\Xi^{0}})(E_{\Xi^{0}} - M_{\Xi^{0}}) + (M_{P_b^{0(2)}} - M_{\Xi^{0}})^2 + 2(M_{\Upsilon})^2]
 \end{align}
 with coupling constant $\overline{g}_{{\Xi^{0}}\Upsilon}$:
 \begin{equation}
\overline{g}_{{\Xi^{0}}\Upsilon} = \frac{g_1}{(2M_{\Xi^0})}
 \end{equation}
 Further, the partial widths expression for triply strange pentaquarks ($P_b^{1^{'}(0)}$) are written as:
  \begin{align}
     \Gamma(P_b^{1^{'}(0)} \rightarrow \Lambda^{0} \Upsilon) = \frac{\overline{g}_{{\Lambda^{0}}\Upsilon}^2}{12\pi}\frac{p_{\Lambda^{0}}}{M_{P_b^{1^{'}(0)}}}(E_{\Lambda^{0}} + M_{\Lambda^{0}})
[2(E_{\Lambda^{0}})(E_{\Lambda^{0}} - M_{\Lambda^{0}}) + (M_{P_b^{1^{'}(0)}} - M_{\Lambda^{0}})^2 + 2(M_{\Upsilon})^2]
 \end{align}
and $\overline{g}_{{\Lambda^{0}}\Upsilon}$:
 \begin{equation}
     \overline{g}_{{\Lambda^{0}}\Upsilon} = \frac{g_1}{(2M_{\Lambda^{0}})}
 \end{equation}
 \\
Expressions for partial widths for the octet particles are listed in Table \ref{tab:decay} in addition to their strong decay channels. Unfortunately, the branching ratios of these decays are still unknown and we have to restrict ourselves to writing partial widths using coupling constant $g_1$ only. However, several works have been proposed about the decay width assignments of hidden-bottom pentaquarks \cite{decay1, decay2, decay3}, by which the coupling constant value may be approximated. We are relying on the experimental observation of hidden-bottom pentaquarks as a substitute for partial widths expressed using the coupling constant, as experimental observation of these particles is currently unavailable.
\begin{table}[ht]
       \centering
        \caption{Expressions for the partial decay width for ${p}\Upsilon$,  ${\Sigma^{+}}\Upsilon$, ${\Xi^{0}}\Upsilon$ and ${\Lambda^{0}}\Upsilon$ channels using the effective Lagrangian method.}
       \begin{tabular}{ccc}
       \hline
       \hline
         Initial state & \hspace{0.2cm} Channel & \hspace{0.3cm} Partial Width[MeV]  \\
           \hline
      $P_b^{(0)}$, $P_b^{+(0)}$ & \hspace{0.2cm} ${p}\Upsilon$ & 752.83$g_1^2$ \\
      \hline
         $P_b^{+(1)}$, $P_b^{0(1)}$, $P_b^{-(1)}$ & \hspace{0.2cm} ${\Sigma^{+}}\Upsilon$ & 558.57$g_1^2$\\
         \hline
         $P_b^{0(2)}$, $P_b^{-(2)}$ & \hspace{0.2cm} ${\Xi^{0}}\Upsilon$ & 496.32$g_1^2$ \\
         \hline
        $P_b^{1^{'}(0)}$ & \hspace{0.2cm} ${\Lambda^{0}}\Upsilon$ & 603.03$g_1^2$ \\
        \hline
        \hline
       \end{tabular}
      \label{tab:decay}
        \end{table}

\section{Summary}
This work consists of the systematic analysis of masses, magnetic moments, and partial widths of hidden-bottom pentaquarks. Firstly, we studied the classification scheme for hidden-bottom pentaquarks using the SU(3) flavor and SU(2) spin representation and assigned to the allowed multiplets. The SU(3) flavor symmetry makes specific predictions about the masses and characteristics of exotic baryons in the octet and decuplet configurations.  In the case of hidden-bottom pentaquarks, singlet, octets, and decuplet are the allowed SU(3) flavor multiplets. We have studied the octet and decuplet configurations with $J^P$ values as $3/2^-$ and $5/2^-$ respectively. We employed effective mass schemes and an extended version of the GR mass formula to compute the masses of hidden-bottom pentaquarks. In recent times, the extension of the Gursey-Radicati mass formula seems to be a useful tool for studying the masses of exotic hadrons. The comparison of our obtained masses using the extension of the Gursey-Radicati mass formula and effective mass scheme for both octet and decuplet demonstrates reasonable consistency with known theoretical data which is shown in Tables  \ref{tab:3} and  \ref{tab:4} respectively. Further, to study the magnetic moments for the ground state hidden-bottom pentaquarks for the octet and decuplet configuration, we used the effective mass and screened charge scheme which demonstrates reasonable consistency with known theoretical data and reported in Tables \ref{tab: magnetic moments of octet} and \ref{tab: magnetic moments} for octet and decuplet configurations respectively. Future experiment studies could explore the possibility of observing a hidden-bottom pentaquark. The $P_c$ pentaquark states were found during the $\Lambda_b$ decay at LHCb. However, there is no such decay for the hidden-bottom pentaquarks. Pentaquarks, made up of three light quarks and a pair of heavy quarks, can be created by stimulating the nucleon with a photon or pion meson, causing a hidden-bottom quark pair to emerge. Thus, in the future, we can see the possibility of hidden-bottom pentaquarks by exciting the nucleon with photons or mesons. Further, by considering the effective Lagrangian approach, we estimated the partial decay widths for the hidden-bottom octet of pentaquarks having $J^P = 3/2^-$, which are reported in Table \ref{tab:decay} with regard to the coupling constant $g_1$ because the branching ratio for these decays is not present and we are relying on the experimental observation of hidden-bottom pentaquarks as a substitute for decay widths expressed in terms of the coupling constant. The study of decay widths contributes to a better understanding of resonances, their internal structure, and interactions. The study of decay width also provides information about the lifetime of a particle. In conclusion, examining decay widths in particle physics is essential for understanding the fundamental features of particles. Thus, our analysis for masses, magnetic moments, and partial widths could be beneficial for upcoming experimental investigations on the hidden-bottom pentaquarks.

%
% once the .bbl file has been generated then place the text in your article.

%This is added by T. Yoneya (editor-in-chief) on 2020/07/09.

\let\doi\relax

%without this code before the command "\begin{thebibliography}{}" , an error will be %flagged. When the bibliography is provided as separate .bib file, then this code %should be placed above the commands "\bibliographystyle{}" and "\bibliography{}" %inside the main TeX file. 
\bibliographystyle{unsrt}
\bibliography{HB} 

\end{document}